\documentclass[aps,pre,twocolumn,superscriptaddress,showpacs,floatfix,longbibliography]{revtex4-1}

\makeatletter
\def\@bibdataout@aps{%
\immediate\write\@bibdataout{%
@CONTROL{%
apsrev41Control%
\longbibliography@sw{%
    ,author="08",editor="1",pages="1",title="0",year="1"%
    }{%
    ,author="08",editor="1",pages="1",title="",year="1"%
    }%
  }%
}%
\if@filesw \immediate \write \@auxout {\string \citation {apsrev41Control}}\fi
}
\makeatother

\usepackage{graphicx}
\usepackage{dcolumn}
\usepackage{bm}

\usepackage{multirow}

\usepackage{amssymb}
\usepackage{amsmath}
\usepackage{graphicx}
\usepackage{xcolor}
\usepackage[colorlinks,citecolor=blue,linkcolor=red,urlcolor=blue]{hyperref}
\usepackage{CJK}
\usepackage{indentfirst}
\usepackage{amsmath}
\usepackage{cases}

\begin{document}
\title{Imaginary-time Quantum Relaxation Critical Dynamics with Semi-ordered Initial States}
\author{Zhi-Xuan Li}
\affiliation{School of Physics and Materials Science, Guangzhou University, Guangzhou 510006, China}

\author{Shuai Yin}
\affiliation{School of Physics, Sun Yat-Sen University, Guangzhou 510275, China}

\author{Yu-Rong Shu}
\email{yrshu@gzhu.edu.cn}
\affiliation{School of Physics and Materials Science, Guangzhou University, Guangzhou 510006, China}
\date{\today}

\begin{abstract}
  We explore the imaginary-time relaxation dynamics near quantum critical points with semi-ordered initial states. Different from the case with homogeneous ordered initial states, in which the order parameter $M$ decays homogeneously as $M\propto \tau^{-\beta/\nu z}$, here $M$ depends on the location $x$, showing rich scaling behaviors. Similar to the classical relaxation dynamics with an initial domain wall in Model A, which describes the purely dissipative dynamics, here as the imaginary time evolves, the domain wall expands into an interfacial region with growing size. In the interfacial region, the local order parameter decays as $M\propto \tau^{-\beta_1/\nu z}$, with $\beta_1$ being an additional dynamic critical exponent. Far away from the interfacial region the local order parameter decays as $M\propto \tau^{-\beta/\nu z}$ in the short-time stage, then crosses over to the scaling behavior of $M\propto \tau^{-\beta_1/\nu z}$ when the location $x$ is absorbed in the interfacial region. A full scaling form characterizing these scaling properties is developed. The quantum Ising model in both one and two dimensions are taken as examples to verify the scaling theory. In addition, we find that for the quantum Ising model the scaling function is an analytical function and $\beta_1$ is not an independent exponent.
\end{abstract}

\maketitle


The understanding of nonequilibrium dynamics in quantum many-body systems is attracting increasing attentions in recent years inspired by the fast developments of quantum computers~\cite{Satzinger2021science,Semeghini2021science,King2022} and ultra-cold atom experimental techniques~\cite{bloch08}. In particular, the study of imaginary-time quantum critical dynamics is heating up thanks to the recent progresses made by quantum computers in realizing imaginary-time evolution in quantum systems~\cite{Motta2020nat,Nishi2021njp}.
Grasping critical properties on the imaginary-time relaxation avenue down to the critical ground state not only saves computational efforts but also yields dynamical properties that is beyond the reach of equilibrium studies.
The critical relaxation dynamics is one of the most simple but important member of the nonequilibrium quantum critical dynamics family. In the past few years, a scaling theory for the imaginary-time quantum critical relaxation dynamics has been developed~\cite{Yin2014prb,Zhangsy2014pre} in analogy to the critical relaxation dynamics in classical systems~\cite{Janssen1989,Lizb1995prl,Lizb1996pre,Zhengb1998ijmpb,Yinghp1998mplb,Zhengb1996prl}.
It has been shown that the initial information can affect the critical relaxation dynamics in the macroscopic time scale owing to the divergence of the correlation time at the critical point.
Universal behaviors have been found during the imaginary-time relaxation process with a homogeneous initial state.
For saturated ordered initial state, the order parameter $M$ changes with the imaginary-time $\tau$ as $M\propto\tau^{-\beta/\nu z}$~\cite{Yin2014prb}, in which $\beta$ is the order parameter exponent defined as $M\propto |g|^{-\beta}$ with $g$ being the distance to the critical point, $\nu$ is the correlation length exponent defined as $\xi\propto |g|^{-\nu}$ with $\xi$ being the correlation length, and $z$ is the dynamic exponent defined as $\zeta\propto \xi^z$ with $\zeta$ being the correlation time.
For initial state with a small initial order parameter $M_0$, the order parameter increases according to $M\propto M_0\tau^{\theta}$ with $\theta$ the critical initial slip exponent in the short-time stage, then decays as $M\propto\tau^{-\beta/\nu z}$ in the long-time stage.
For initial state with arbitrary order parameter, a universal characteristic function is introduced to describe the universal effects induced by the initial state~\cite{Zhengb1996prl,Zhangsy2014pre}.
These phenomena have been investigated in various phase transitions within and beyond the Landau paradigm~\cite{Yin2014prb,Zhangsy2014pre,Shu2017prb,Shu2020prb,Shu2021prl,Shu2022prb}.
It has been shown that the imaginary-time relaxation dynamics of quantum systems can be different from its classical counterpart. For instance, the critical initial slip exponent $\theta$ is $0.373$~\cite{Yin2014prb} and $0.209$~\cite{Shu2017prb} for the one-dimensional ($1$D) and two-dimensional ($2$D) quantum Ising model, while $\theta$ is $0.191$~\cite{okano1997npb} and $0.108$~\cite{Jaster1999jpa} for their classical counterparts, respectively. 
Therefore, the imaginary-time relaxation dynamics deserves special attentions.
The flourishing developments in this issue also inspire us to explore the effects induced by other kinds of initial states, like the inhomogeneous initial state, in the imaginary-time evolution.

Critical properties in the presence of the inhomogeneous interfacial regions have raised long-term attentions in various systems, since phase coexistence is a common phenomenon in nature~\cite{Dombbook}.
In particular, critical relaxation dynamics with a domain interface in a semi-ordered initial state was studied~\cite{Zhou2007epl,Zhou2008pre}.
These works showed that different from the relaxation dynamics with a homogeneous initial state, the initial domain wall can expand into a growing interfacial region, and in this region the order parameter decays obeying a distinct scaling relation $M\propto t^{-\beta_1/\nu z}$ with $\beta_1$ being an additional dynamic exponent.
In quantum systems, exotic prethermal dynamics induced by the interface in the $2$D quantum Ising model was discovered~\cite{Atsuki2022prl,Oliver2022,Federico2022}.

\begin{figure}[!htbp]
  \centering
  \includegraphics[width=0.75\linewidth,clip]{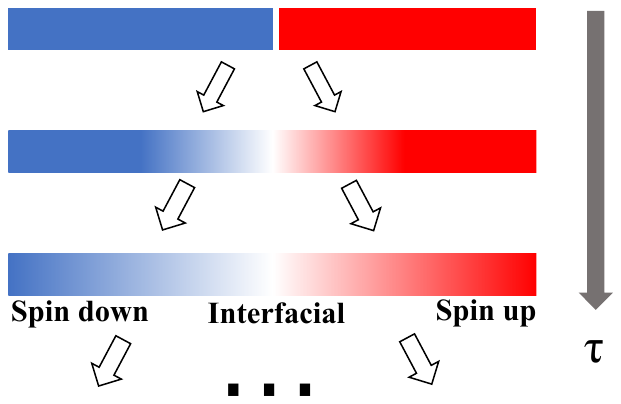}
  \caption{Sketch of the imaginary-time relaxation dynamics from a semi-ordered initial state. A sharp domain wall separates the spin-up and -down domains in the initial state. As time evolves, the domain wall extends to an interfacial region.}
  \label{fig:illu}
\end{figure}

Motivated by the above intriguing issues, we investigate the imaginary-time relaxation dynamics with a semi-ordered initial state, in which two completely ordered domains with opposite spin direction sandwich a sharp domain wall, as shown in Fig.~\ref{fig:illu}. We find that similar to the classical case~\cite{Zhou2007epl,Zhou2008pre},
as the time evolves, the sharp domain wall will blur and expand into an interfacial region.
Let us focus on the behavior of the local parameter at the position $x$, with $x$ denotes the distance to the initial domain wall.
When $x$ is far away from the interfacial region, the local order parameter decays as $M(\tau,x)\propto \tau^{-\beta/\nu z}$. As time elapses, the interfacial region spreads to the position $x$.
Accordingly, the order parameter changes to decay as $M(\tau,x)\propto \tau^{-\beta_1/\nu z}$. Here, $\beta_1$ is a purely dynamic exponent, since it has no equilibrium counterpart, similar to the classical case~\cite{Zhou2007epl,Zhou2008pre}. A full scaling form is then developed to explain this behavior.
We take the $1$D and $2$D quantum Ising models as examples to verify this scaling theory.
From the numerical results, we find that the scaling function is an analytical function and $\beta_1$ seems not an independent critical exponent. Instead, it satisfies $\beta_1/\nu z=\beta/\nu z+1$, in analogy to the classical case~\cite{Zhou2007epl,Zhou2008pre}.

For the imaginary-time relaxation dynamics, the evolution of the wave function $|\psi(\tau)\rangle$ obeys the imaginary-time Schr\"{o}dinger equation $-\partial_\tau|\psi(\tau)\rangle=H|\psi(\tau)\rangle$
with the normalization condition $\langle\psi(\tau)|\psi(\tau)\rangle=1$~\cite{Yins2014prb,Yins2014pre}.
The formal solution of the Schr\"odinger equation is given by $|\psi(\tau)\rangle=\exp(-\tau H)|\psi(\tau_0)\rangle/Z$, in which $\tau_0$ is the initial time of the evolution and $Z=\Vert \exp(-\tau H)|\psi(\tau_0)\rangle \Vert$, with $\Vert \cdot \Vert$ being the modulo operation.
Since both the imaginary-time dynamics and the classical dynamics described by Model A, which refers to a set of models with no conservation laws~\cite{Hohenberg1977rmp}, are purely dissipative dynamics, one expects that their dynamic scaling behaviors near the critical point are similar~\cite{Yin2014prb}.
For instance, from the saturated ordered initial state, the order parameter $M$ follows the same power-law decay with the evolution time with the exponent $\beta/\nu z$ for both classical model A dynamics and quantum imaginary-time relaxation dynamics.

Here we study the influence of a different type of initial state to the quantum imaginary-time relaxation dynamics. The initial state is set as a semi-ordered state with two fully-ordered domains with opposite spin directions, separated by a sharp domain wall, as shown in Fig.~\ref{fig:illu}. Since the translational symmetry is broken by the initial state, it is expected that the evolution behavior of $M$ depends on the distance to the initial domain wall $x$.
Near the domain wall, the spin at small $x$ feels a stronger spin-flip intention from the other domain where the spins orientated in the opposite direction than its homogeneous ordered environment. 
Accordingly, one anticipates that at the critical point, for small $x$, the local order parameter should follows
\begin{equation}
  \label{msmallx}
  M(x,\tau)\propto \tau^{-\beta_1/\nu z}, 
\end{equation}
with $\beta_1$ being an additional critical exponent which is larger than $\beta$, since the other domain lures the spin at $x$ to flip.
Note that similar to the classical case~\cite{Zhou2007epl,Zhou2008pre}, here $\beta_1$ is a purely nonequilibrium critical exponent, since it has no static counterpart.

In contrast, for large $x$, the dynamic scaling behavior of $M(x,\tau)$ is much richer as a result of the spread of the effects induced by the domain wall, as illustrated in Fig.~\ref{fig:illu}. In the short-time stage, the domain wall region is too far away to control the spin at $x$ 
and thus the local order parameter $M$ decays according to $M(x,\tau)\propto \tau^{-\beta/\nu z}$, similar to the case with homogeneous ordered initial state. Therefore, this stage is referred to as the `{\it homogeneous region}'. As time passes by, the domain wall extends into an `{\it interfacial region}' with growing size. When the location $x$ is absorbed into this interfacial region, $M$ will evolve following Eq.~(\ref{msmallx}).

\begin{figure}[!htbp]
  \centering
  \includegraphics[width=\linewidth,clip]{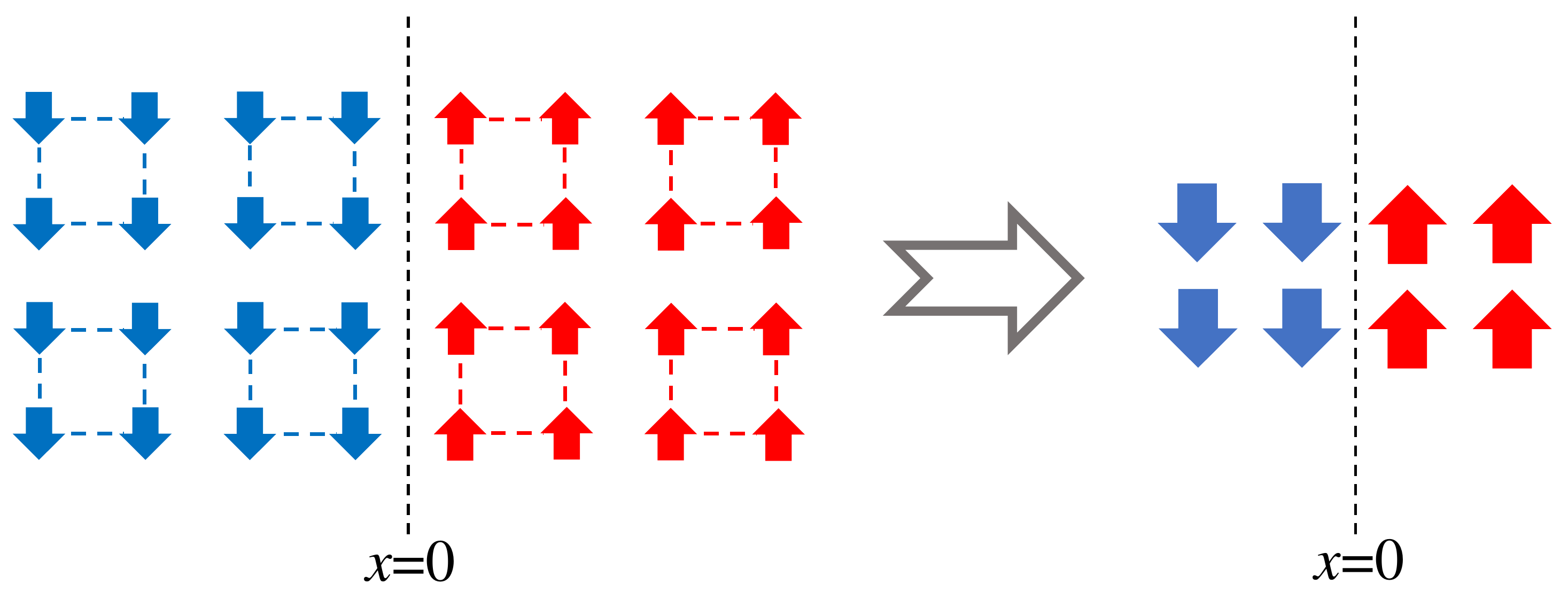}
  \caption{Sketch of renormalization group transformation of the initial state. After the processes of coarse graining and rescaling, the state keeps invariant.
  }
  \label{fig:illu2}
\end{figure}

{\it Scaling theory.} For the relaxation critical dynamics with homogeneous initial state, a characteristic scaling behavior is the appearance of the critical initial slip characterized by an independent critical exponent $\theta$. A nature question is whether or not $\beta_1$ is another independent exponent. To answer this question, we develop a full scaling form to describe the whole imaginary-time relaxation process with the semi-ordered initial state. In analogy with the classical situation~\cite{Zhou2007epl,Zhou2008pre}, the scaling form of the local order parameter $M$ at a quantum critical point is given by
\begin{equation}
  \label{scalingtr}
  M(\tau,x)=\tau^{-\beta/\nu z}f(x \tau^{-1/z}),
\end{equation}
in which $f(x \tau^{-1/z})$ is the scaling function. By comparing Eq.~(\ref{scalingtr}) with the imaginary-time relaxation scaling theory with the homogeneous initial state~\cite{Yin2014prb}, one finds that the initial order parameter is absent in Eq.~(\ref{scalingtr}). The reason is that the initial state keeps invariant under the renormalization transformation, as illustrated in Fig.~\ref{fig:illu2}. Similarly, the initial correlation is also not included since both the initial correlation length and correlation time are zero.

It is expected that the scaling behaviors discussed above should be covered by the full scaling form Eq.~(\ref{scalingtr}).
This gives some constraints on the scaling function $f(x \tau^{-1/z})$: (i) $f(x \tau^{-1/z})$ should be an odd function of $x$ since $M$ should change sign on switching the spin orientation of the initial spin domains; (ii) in the homogeneous region with large $x$ and small $\tau$, $M(x,\tau)$ should satisfy $M(x,\tau)\propto \tau^{-\beta/\nu z}$; (iii) in the interfacial region with small $x$ and large $\tau$, $M(x,\tau)$ should decay as $M(x,\tau)\propto\tau^{-\beta_1/\nu z}$.
Thus, the scaling function $f(x \tau^{-1/z})$ must tend to a constant as $x \tau^{-1/z}\rightarrow\infty$, while for $x \tau^{-1/z}\rightarrow 0$, $f(x \tau^{-1/z})$ should satisfy $f(x \tau^{-1/z})\propto |x\tau^{-1/z}|^\omega$, in which $\omega\equiv(\beta_1-\beta)/\nu$. Accordingly, we summarize the properties of $f(x \tau^{-1/z})$ as follows,
\begin{subnumcases}
  {f=}
  C_1&,  $x \tau^{-1/z}\gg1$ \label{scalingfa}\\
  C_2 {\rm sgn}(x)(x \tau^{-1/z})^{\omega}&, $x \tau^{-1/z}\ll1$ \label{scalingfb}
\end{subnumcases}
in which both $C_1$ and $C_2$ are constants, and ${\rm sgn}(x)$ denotes the sign function of $x$. In general, without extra information, $\omega$ cannot be determined from the phenomenological scaling analyses. In this respect, $\beta_1$ can be an independent additional exponent. However, for the special case in which $f(x \tau^{-1/z})$ is an analytical function, the Taylor expansion of $f(x\tau^{-1/z})$ at $x \tau^{-1/z}\rightarrow 0$, gives $\omega=1$, which indicates that $\beta_1$ is not an independent exponent but instead satisfies $\beta_1/\nu=\beta/\nu+1$. For finite-size system with $L$ its linear size, by taking into account the finite-size effects, the full scaling form reads
\begin{equation}
  \label{scalingtrL}
  M(\tau,x,L)=\tau^{-\beta/\nu z}f_L(x \tau^{-1/z},L\tau^{-1/z}),
\end{equation}
in which $f_L$ is another scaling function. When $L\tau^{-1/z}\gg 1$, the interfacial region is much smaller than the lattice size. Accordingly, the finite-size effects can be ignored and $f_L(x \tau^{-1/z},L\tau^{-1/z})$ can be approximated as $f(x \tau^{-1/z})$. In contrast, when $L\tau^{-1/z}\ll 1$, the energy gap induced by the lattice size becomes relevant, the system will decay exponentially towards the ground state. This is the usual finite-size scaling region.

{\it Model.} The Hamiltonian of the quantum Ising model studied here is ~\cite{Sachdevbook}
\begin{equation}
  \label{eq:hamiltonian}
  H=-J\sum_{\langle ij\rangle}\sigma^z_i\sigma^z_j-h_x\sum_{i}\sigma^x_i,  
\end{equation}
in which $J$ is set as $1$ as the unity of energy scale, and $h_x$ is the strength of the transverse field. $\sigma^{z,x}_i$ denotes the Pauli matrix in $z,x$-direction at site $i$, and $\langle ij\rangle$ represents nearest neighbors. For large $h_x$, the system is in the quantum paramagnetic state, and for small $h_x$, the system is in the ferromagnetic state. For the $1$D case, the critical point locates at $h_{x}=h_{xc}=1$, and the critical exponents are exactly solved as $\beta=1/8$, $\nu=1$, and $z=1$~\cite{Sachdevbook,Sondhi1997rmp,Vojta2003rpp}. For the $2$D case, the critical point is at $h_x=h_{xc}=3.04451$~\cite{Shu2017prb}. The critical exponents are estimated as $\beta=0.327(1)$, $\nu=0.630(2)$~\cite{Jaster1999jpa}, and $z=1$~\cite{Sachdevbook,Sondhi1997rmp,Vojta2003rpp}.

\begin{figure}[!htbp]
  \centering
  \includegraphics[width=\linewidth,clip]{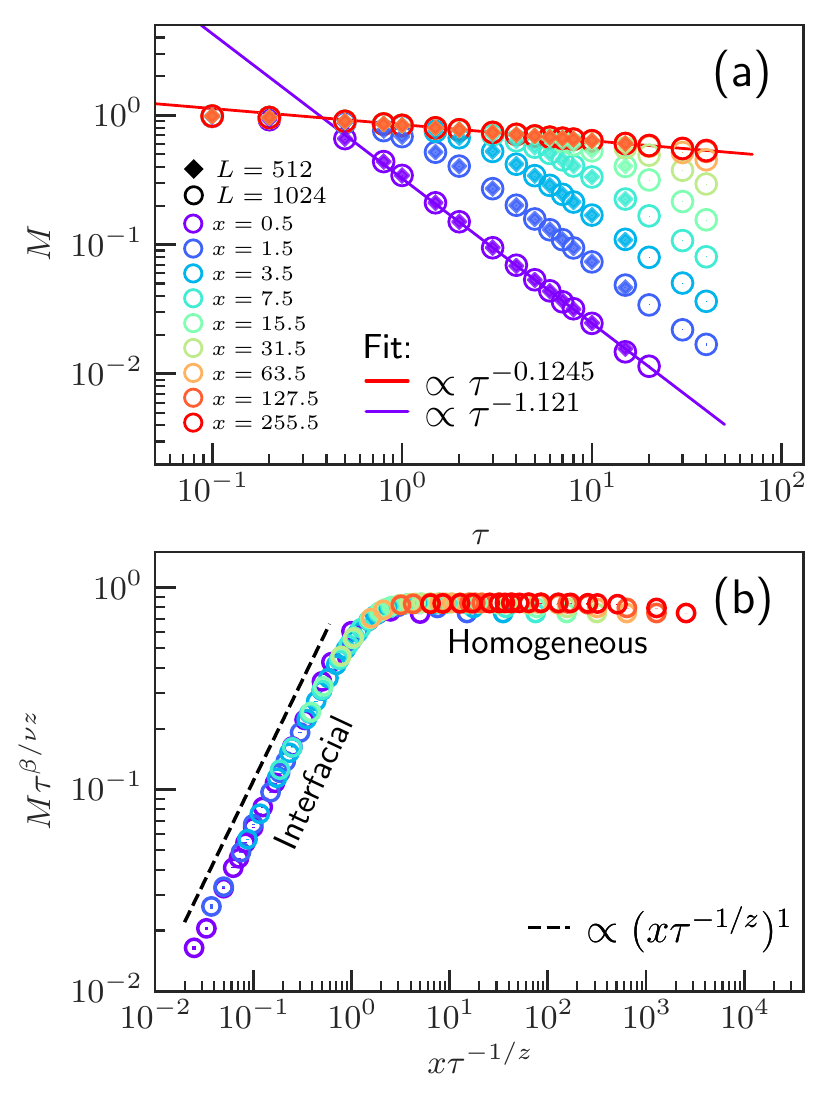}
  \caption{Local order parameter $M(x,\tau)$ for the $1$D quantum Ising model. Log-log scale is used. Panel (a): Solid diamonds and open circles represent results of $L=512$ and $1024$, respectively. Crossover behaviors are shown in $M(x,\tau)$ for different $x$. Solid lines are power-law fits to the data of $x=0.5$ and $x=255.5$. The critical exponent $\beta/\nu$ and $\beta_1/\nu$ are given by $0.1245(5)$ and $1.121(9)$, respectively. Panel (b): The rescaled curve shows two different scaling region and the dash line represents a power law of $(x\tau^{-1/z})^{\omega}$ with $\omega=1$.}
  \label{fig:infinit1024}
\end{figure}

To verify the above scaling theory, we perform projector quantum Monte Carlo~\cite{sandvik2010review,Farhi2012prb} simulations in both the $1$D and $2$D quantum Ising model.
We compute dependence of the local order parameter on the position $x$ and the evolution time $\tau$. For simplicity, the periodic boundary condition is used so that boundary effects can be neglected.
At a given evolution time $\tau$, for $1$D system, the local order parameter is defined as $M(x)=\langle\sigma^z(x)\rangle$.
In the $2$D system, since the $y$ direction is homogeneous, $M(x)$ is taken as the $y$-direction averaged value for a given $x$, defined as $M(x)=\left\langle\sum_{y=1}^{L}\sigma^z(x)\right\rangle/L$,
in which $y$ direction is perpendicular to the $x$ direction.
The lattice shape is set to be $2L\times L$ so that the two domains occupy the same area of $L\times L$. Note that in Eqs.~(\ref{scalingtr})-(\ref{scalingtrL}), $x$ is a continuous variable that represents the distance to the center of the interfacial region, but here in lattice system, $x$ is discrete. 
As shown in Fig.~\ref{fig:illu2}, the location of the initial domain wall is set as $x=0$ and for other positions, $x$ should be among $\{\pm\frac{1}{2},\pm\frac{3}{2},...,\pm(\frac{L_x}{4}+\frac{1}{2})\}$ with $L_x=L$ for $1$D and $L_x=2L$ for $2$D, restricted by the periodic boundary condition.
Besides, due to the spin inversion symmetry, $M(x)$ and $-M(-x)$ should be equivalent, so that the finial result $M(x)$ is taken as the average of $|M(x)|$ and $|M(-x)|$.

{\it Numerical results.} For the $1$D quantum Ising model, in Fig.~\ref{fig:infinit1024} (a) we show the evolution of the order parameter for different $x$. In order to extract the exponent without the influence of finite-size effect, we plot $M$ for system of $L=512$ and $1024$ find that up to the time scale considered here, the dependence of $M$ on $L$ is very weak. Therefore, one can consider the fitted result represents the effects induced by the initial domain wall only. At the position $x=0.5$ that is closest to the initial domain wall, we find that after a transient non-universal stage, the order parameter presents a power-law decay of $M\propto \tau^{-1.121}$. The exponent $1.121(9)$ (with the number in the parentheses stands for error bar) is much larger than $\beta/\nu z=1/8$ characterizing the evolution of $M$ relaxed from the homogeneous initial state.
As the scaling theory indicates, close to the initial domain wall, the local order parameter should decay with an exponent $\beta_1/\nu z$, rather than $\beta/\nu z$. With $\nu=1$ and $z=1$ known, one deduces that $\beta_1=1.121(9)$ and one can find $\omega=0.9965$. In consideration of the statistical error, these results strongly indicate that $\beta_1=9/8$ and $\omega=1$, suggesting that the scaling function $f(x\tau^{-1/z})$ is an analytical function.

\begin{figure}[!htbp]
  \centering
  \includegraphics[width=\linewidth,clip]{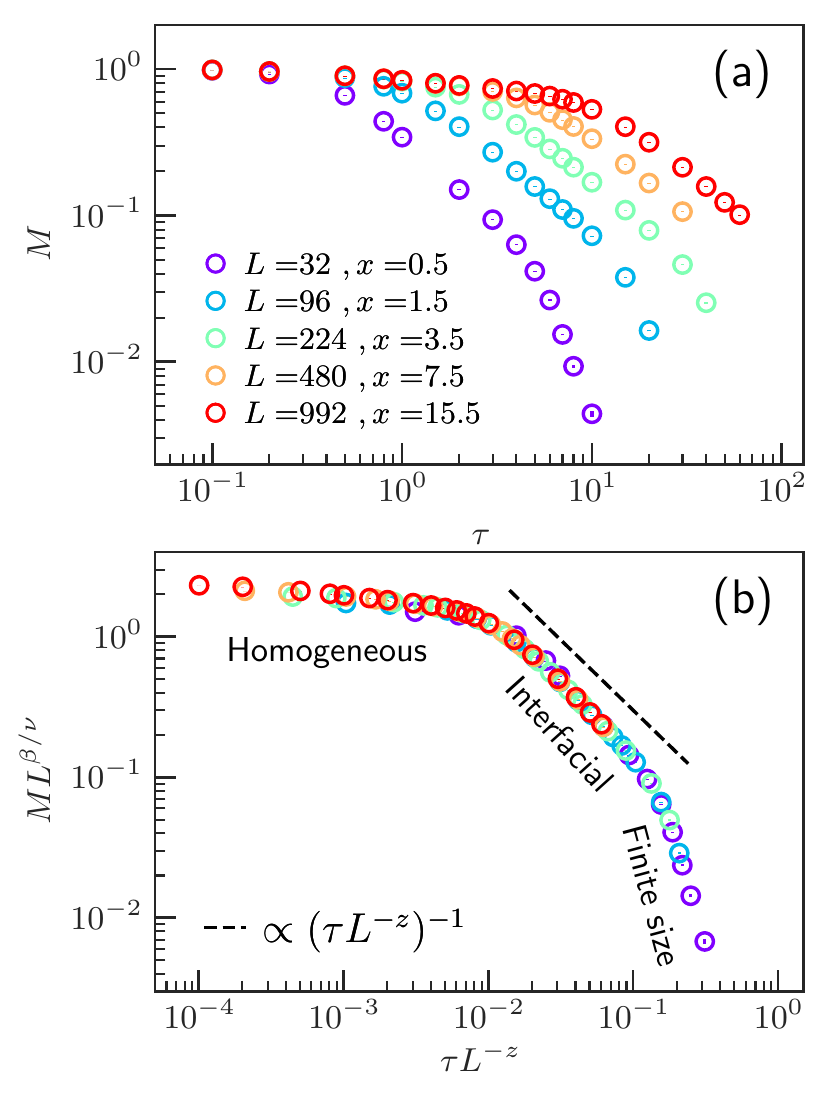}
  \caption{For fixed $xL^{-1}=1/64$, evolution of $M$ before (a) and after (b) rescaling for $L=32$ to $992$ in $1$D quantum Ising model. Three different scaling regions are labeled in (b). The dash line is a power law of $(\tau L^{-z})^{-\omega}$ with $\omega=1$. Log-log scale is used.}
  \label{fig:finite_scale}
\end{figure}

Different from the results of $x=0.5$, at positions far away from the initial domain wall, for instance, $x=255.5$ for $L=1024$, the local order parameter decays as $M\propto\tau^{-0.1245}$ in the universal short-time stage, as shown in Fig.~\ref{fig:infinit1024} (a). The exponent $0.1245(5)$ is close to $\beta/\nu z$, demonstrating that in the short-time stage, the evolution time is yet too short for the influence of the other domain to propagate to the positions far away from the initial domain wall. Thus, the spin at large $x$ only feels its own homogeneous background, where the homogeneous fluctuations dominate.
As time elapses, $M$ gradually crosses over to the behavior of $M\propto \tau^{-\beta_1/\nu z}$, indicating that the influence induced by the initial inhomogenity begin to control the dynamics when the interfacial region spreads over $x$. This crossover behavior can be clearly observed for intermediate $x$. The crossover time scale depends on the position $x$.

In Fig.~\ref{fig:infinit1024} (b), we rescale $M$ and $\tau$ as $M\tau^{\beta/\nu z}$ and $x \tau^{-1/z}$, respectively. We find that the rescaled curves collapse on a single curve, confirming Eq.~(\ref{scalingtr}). This single curve is just the scaling function $f(x \tau^{-1/z})$. For $x \tau^{-1/z}\gg 1$ the local order parameter is in the homogeneous region and $f(x\tau^{-1/z})$ tends a constant, while for $x \tau^{-1/z}\ll 1$ the local order parameter is in the interfacial region and $f(x \tau^{-1/z})\propto (x \tau^{-1/z})^{\omega}$ with $\omega$ close to $1$. These results not only verify Eqs.~(\ref{scalingfa}) and (\ref{scalingfb}), but also indicate that the scaling function $f(x\tau^{-1/z})$ is an analytical function for $x \tau^{-1/z}\rightarrow 0$.

\begin{figure}[!htbp]
  \centering
  \includegraphics[width=\linewidth,clip]{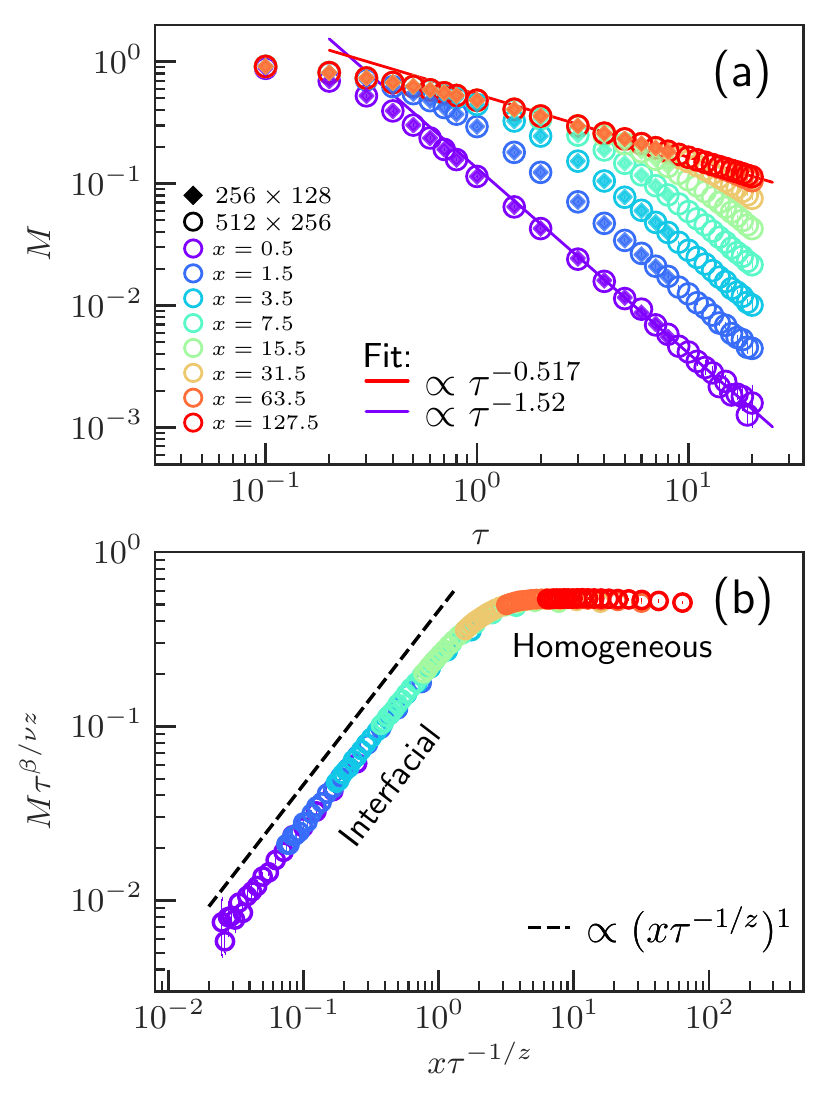}
  \caption{Local order parameter $M(x,\tau)$ for the $2$D quantum Ising model. Log-log scale is used. Panel (a): Solid diamonds and open circles represent results of lattice size of $256\times 128$ and $512\times 256$. Finite-size effects are very weak. Crossover behaviors are seen when $x$ changes. The solid lines are power-law fits to the data of $x=0.5$ and $x=127.5$. The critical exponent $\beta/\nu$ and $\beta_1/\nu$ are given by $0.517(4)$ and $1.52(6)$, respectively.
  In panel (b): Rescaled curve of $M$. Different scaling regions are labeled and the dash line represents a power law of $(x\tau^{-1/z})^{\omega}$ with $\omega=1$.}    
  \label{fig:infinit2D}
\end{figure}

Then we explore the finite-size effects.
For convenience, a variable transformation can be implemented in Eq.~(\ref{scalingtrL}) by replacing $x\tau^{-1/z}$ and $L\tau^{-1/z}$ with $\tau L^{-z}$ and $xL^{-1}$, respectively, giving the scaling form $M(\tau,x,L)=L^{-\beta/\nu}g(\tau L^{-z},xL^{-1})$, with $g$ being another scaling function.
For fixed $xL^{-1}$, the curves of $M$ for different lattice sizes are shown in Fig.~\ref{fig:finite_scale} (a). After rescaling $M$ and $\tau$ as $M L^{\beta/\nu}$ and $\tau L^{-z}$, respectively, we find in Fig.~(\ref{fig:finite_scale}) (b) the curves collapse on to the single curve of $g(\tau L^{-z})$, confirming Eq.~(\ref{scalingtrL}).
With $xL^{-1}$ fixed, for small $\tau L^{-z}$, $g(\tau L^{-z})$ tends to a constant, corresponding to the homogeneous region. For intermediate $\tau L^{-z}$, $g(\tau L^{-z})\propto (\tau L^{-z})^{-1}$, which is the interfacial region. For large $\tau L^{-z}$, there exists a finite-size region in $g(\tau L^{-z})$, where finite-size effect starts to take control of the scaling function. In the finite-size region, $M$ decays very fast towards its equilibrium value. The good collapse again verifies the scaling theory proposed above.
  
For the $2$D quantum Ising model, in Fig.~\ref{fig:infinit2D} (a) we show the evolution of the local order parameter for different $x$.
Similarly, here we plot the results of two different system sizes $256\times 128$ and $512\times 256$, in order to show finite-size effects are very weak at the time scale considered.
For spins closest to the initial domain wall, namely $x=0.5$, Fig.~\ref{fig:infinit2D} (a) shows that the local order parameter decays as $M\propto \tau^{-1.52}$ with the exponent $1.52(6)$ that is apparently larger than $\beta/\nu=0.518$~\cite{Shu2017prb}, suggesting that the spins at small $x$ are strongly affected by the initial inhomogenity. Similar to the $1$D case, at $x=0.5$, $M$ should decay as $\tau^{-\beta_1/\nu z}$. Therefore, one finds that $\beta_1/\nu z=1.52(6)$.
Differently, for spins far away from the initial domain wall with large $x$, in the short-time stage, the local order parameter exhibits a slower power-law decay as $M\propto \tau^{-0.517}$ for the largest value $x=127.5$. The exponent $0.517(4)$ is close to $\beta/\nu z=0.518$~\cite{Shu2017prb}. As time passes, $M$ gradually crosses over to a power-law decay with the exponent $\beta_1/\nu z$, indicating that the interfacial region spreads across $x$ . The crossover time scale increases with $x$ moving away from the initial domain wall. From the results of $x=0.5$ and $x=127.5$, we obtain $\omega=1.003$, which is also close to $1$. In analogy to the $1$D case, taking into account of statistical errors and the possible remaining finite-size effects, one deduces that $\omega=1$ for the $2$D case as well.

In Fig.~\ref{fig:infinit2D} (b), we rescale $M$ and $\tau$ as $M\tau^{\beta/\nu z}$ and $x\tau^{-1/z}$. The rescaled curves collapse onto a single curve of scaling function $f(x\tau^{-1/z})$, confirming Eq.~(\ref{scalingtr}).
Similar to the $1$D case, when $x\tau^{-1/z} \gg 1$, $f(x\tau^{-1/z})$ tends to a constant, representing the homogeneous region. When $x\tau^{-1/z} \ll 1$, the scaling function is proportional to $(x\tau^{-1/z})^{\omega}$, with $\omega=1$, corresponding to the interfacial region. These results verify Eqs.~(\ref{scalingfa}) and (\ref{scalingfb}) and indicate that the scaling function is analytical.

{\it Discussions.} The agreement between the results of the $1$D and $2$D quantum Ising model confirms that the scaling theory discussed above is universal.
We find $\omega$ is very close to $1$, which strongly suggests that the scaling function $f(x\tau^{-1/z})$ is an analytical function when $x\tau^{-1/z}\rightarrow 0$ and $\beta_1$ is not an independent critical exponent.The analyticity of the scaling function $f(x\tau^{-1/z})$ may originated from the preparation of the semi-ordered initial state. For the present case, the spin-up and spin-down domain correspond to the two saturated initial states, which are both fixed points of the scaling transformation. Moreover, although the initial state contributes a sharp domain wall, this non-analyticity can be smeared by the quantum fluctuations in the following relaxation.

Recent experiments have shown that the quantum imaginary-time evolution can be implemented in quantum computers~\cite{Nishi2021njp,Motta2020nat}.
The imaginary-time evolution in a small time interval $\Delta \tau$ can be approximated by the real-time evolution operator with the auxiliary Hamiltonian $\mathcal{H}$ determined by minimizing the residual norm $\left\Vert e^{-\Delta \tau H}|\psi\rangle/Z-e^{-i\Delta t\mathcal{H}}|\psi\rangle\right\Vert ^2$ where $Z=\langle\psi|e^{-2 \Delta \tau H}|\psi\rangle$~\cite{Nishi2021njp,Motta2020nat}.
It is promising that the quantum critical relaxation dynamics can be realized during the process to the ground state in quantum devices. The semi-ordered initial state considered here has vanishing correlation length and is easy to prepare, making the issue stuied here a suitable candidate for realizations of nonequilibrium dynamics in quantum computations.

Here we discuss the relaxation dynamics with semi-ordered initial states of a generalized quantum Ising model, which includes an additional longitudinal field term $-h_z\sum_{i}\sigma_i^z$ to the original model~(\ref{eq:hamiltonian}), described by $H=H_{0}-h_z\sum_{i}\sigma_i^z$. $H_0$ denotes the Hamiltonian in~(\ref{eq:hamiltonian}).
In the thermodynamic limit, the scaling behavior of the local order parameter is
\begin{equation}
  M=\tau^{-\beta/\nu z}f(x\tau^{-1/z},h_z\tau^{\beta\delta/\nu z}),
  \label{eq:hz}
\end{equation}
in which the exponent $\delta$ is defined by $M\propto h_z^{1/\delta}$ at the critical point and $f$ is a scaling function. Imposing a longitudinal field aligned along the spin-up direction in Fig.~\ref{fig:illu}, up to linear term of $f$, one obtains
\begin{equation}
M=\tau^{-\beta/\nu z}f'(0,0)(x\tau^{-1/z}+h_z\tau^{\beta\delta/\nu z}),
\end{equation}
in which $f'$ denotes the first derivative of the scaling function $f$.
To find out the location of the domain wall, one should let $M=0$ since the domain wall separates the spin-up and -down domains. For $h_z=0$, enforcing $M=0$ one finds that $x=0$, as shown in Fig.~\ref{fig:illu2}. For $h_z\neq 0$, enforcing $M=0$ one arrives at $x\tau^{-1/z}+h\tau^{\beta\delta/\nu z}=0$, giving $x=-h_z\tau^{\beta\delta/\nu z+1/z}$. Therefore, in the presence of the longitudinal field, the domain wall moves towards the spin-down side and the stronger $h_z$ is, the further $x$ is. The longitudinal field tends to force all spins align along the spin-up direction homogeneously. As $\tau\rightarrow \infty$, the order parameter $M\propto h_z^{1/\delta}$, recovering the equilibrium scaling behavior. As a result, the relaxation dynamics is a superposition of the two effects: the domain wall should shifts towards the side where spins align opposite to $h_z$ and extends into an interfacial region.

In summary, we study the nonequilibrium imaginary-time quantum critical dynamics with semi-ordered initial states. We show that in the imaginary-time relaxation process the domain wall extends to an interfacial region with growing size. We find that the local order parameters inside and outside of the interfacial region satisfy different scaling relations. When the location is outside of the interfacial region, the order parameter evolves as $M\propto\tau^{-\beta/\nu z}$, similar to the case with homogeneous initial ordered state. In contrast, when the location is inside the interfacial region, the order parameter evolves as $M\propto\tau^{-\beta_1/\nu z}$. By analogy with the classical critical dynamics, we develop a scaling theory and verify it numerically in the $1$D and $2$D quantum Ising model. The numerical results show that scaling theory is universal and strongly indicate that $\beta_1$ is not an independent exponent but satisfies $\beta_1/\nu=\beta/\nu+1$. We also discuss the relaxation dynamics with semi-ordered initial states in the presence of an additional longitudinal field to the quantum Ising model.

{\it Acknowledgements.} We gratefully acknowledge helpful discussions with Bo Zheng. We thank an anonymous referee for his/her suggestions of considering the relaxation dynamics in the presence of a longitudinal field term of the quantum Ising model.
Z.X.L. and Y.R.S. are supported by the National Natural Science Foundation of China (Grants No. 12104109) and the Science and Technology Projects in Guangzhou (202201020222). S.Y. is supported by the Science and Technology Projects in Guangzhou (202102020367) and the Fundamental Research Funds for Central Universities, Sun Yat-Sen University(22qntd3005).

\bibliography{ref1}
\itemsep=-1pt plus.2pt minus.2pt

\end{document}